\documentclass[preprint2]{aastex}

\slugcomment{Not to appear in Nonlearned J., 45.}

\shorttitle{Star Formation Law in Taffy I }
\shortauthors{Komugi and Tateuchi et al.}

\begin{document}


\title{The Schmidt-Kennicutt Law of Matched-Age Star Forming Regions; Pa$\alpha$
Observations of the Early-Phase Interacting Galaxy Taffy I}


\author{S. Komugi\altaffilmark{1}}
\affil{Joint ALMA Observatory, Alonso de Cordova 3107, Vitacura, Santiago 763-0355, Chile}
\email{skomugi@alma.cl}
\author{K. Tateuchi, K. Motohara, N. Kato, M. Konishi, \\ S. Koshida, T. Morokuma, H. Takahashi, T. Tanab\'{e}, Y. Yoshii}
\affil{Institute of Astronomy, The University of Tokyo, Osawa 2-21-1, Mitaka, Tokyo 181-0015, Japan}
\author{T. Takagi}
\affil{Institute of Space and Astronautical Science, JAXA, 3-31-1 Yoshinodai, Sagamihara, Kanagawa 229-8510, Japan}
\author{D. Iono, H. Kaneko\altaffilmark{3,4}}
\affil{Nobeyama Radio Observatory, National Astronomical Observatory, 462-2, Nobeyama, Minamimaki, Minamisaku, Nagano 384-1305, Japan}
\author{J. Ueda\altaffilmark{2}}
\affil{Harvard-Smithsonian Center for Astrophysics, 60 Garden Street, Cambridge, MA02138}
\author{T. R. Saitoh}
\affil{Interactive Research Center of Science, Tokyo Institute of Technology, 2-12-1, Ookayama, Meguro 152-0033, Japan}

\altaffiltext{1}{National Astronomical Observatory of Japan, Osawa 2-21-1, Mitaka, Tokyo 181-8588, Japan}
\altaffiltext{2}{Department of Astronomy, School of Science, The University of Tokyo, Hongo 7-3-1, Bunkyo-ku, Tokyo 113-0033, Japan}
\altaffiltext{3}{Department of Astronomical Science, School of Physical Sciences, The Graduate University of Advanced Studies (SOKENDAI), 2-21-1 Osawa, Mitaka, Tokyo 181-8588, Japan}
\altaffiltext{4}{Institute of Physics, University of Tsukuba, 1-1-1 Tennodai, Tsukuba, Ibaraki}

\begin{abstract}
In order to test a recent hypothesis that the dispersion in the Schmidt-Kennicutt law arises from variations in
 the evolutionary stage of star forming molecular clouds, we compared molecular gas and recent star formation in
 an early-phase merger galaxy pair, Taffy I (UGC\ 12915/UGC\ 12914, VV\ 254) which went through a direct
collision 20 Myr ago and whose star forming regions are expected to have similar ages.  Narrow-band Pa$\alpha$
 image is obtained using the ANIR near-infrared camera on the mini-TAO 1m telescope.  The image enables us to 
derive accurate star formation rates within the galaxy directly.  The total star formation rate, 
$22.2\ M_\odot \mathrm{yr^{-1}}$, was found to be much higher than previous estimates.  Ages of individual star forming
 blobs estimated from equivalent widths indicate that most star forming regions are $\sim$ 7 Myr old, except for a
 giant HII region at the bridge which is much younger.  Comparison between star formation rates and molecular gas masses 
for the regions with the same age exhibits a surprisingly tight correlation, a slope of 
unity, and star formation efficiencies comparable to those of starburst galaxies.  These results suggest that Taffy I
 has just evolved into a starburst system after the collision, and the star forming sites are at a similar stage in
their evolution from natal molecular clouds except for the bridge region.  The tight Schmidt-Kennicutt law supports the
 scenario that dispersion in the star formation law is in large part due to differences in evolutionary stage of star forming regions.
\end{abstract}





\section{Introduction}
Understanding the nature and physics which underly the relation between star formation rate (SFR) and
 gas density is crucial for understanding galaxy evolution and star formation history
 of the universe.  The most thoroughly investigated expression of this relation is the Schmidt-Kennicutt (SK) law
\citep{kennicutt98}, the empirical power-law relation between surface densities of SFR ($\Sigma_\mathrm{SFR}$) and  molecular gas 
($\Sigma_\mathrm{H2}$), written as 
\begin{equation}
\Sigma_\mathrm{SFR} \propto \Sigma_\mathrm{H2}^N.
\end{equation}
  The power law index $N$ is typically found to be $N>1$ for the commonly used molecular gas tracers $^{12}\mathrm{CO}(J=1-0)$
\citep{heyer04, komugi05} and $^{12}\mathrm{CO}(J=2-1)$ \citep{schuster07}, but $N\sim 1.0$ when dense gas tracers such as high CO transitions and HCN
are used to trace $\Sigma_\mathrm{H2}$ \citep{gao04, komugi07, bigiel08, muraoka09, iono09}.  This is typically interpreted as the dense gas
tracing individual units of star formation \citep{wu05}, which forms stars at a constant efficiency.

A less emphasized aspect of the SK law is its dispersion around the relation.  Much of the
 dispersion can be attributed to uncertainties in calibrating $\Sigma_\mathrm{SFR}$ and $\Sigma_\mathrm{H2}$, as 
frequently used massive star formation tracers such as H$\alpha$ are strongly affected by dust extinction 
\citep{komugi07}, and molecular gas mass depends on the assumed CO-to-$\mathrm{H_2}$ conversion factor, 
$\mathrm{X_{CO}}$, which can vary significantly with environment \citep[e.g.,][]{arimoto96, israel97, komugi11}.
  SK laws studied over a scale of several kilo-parsecs \citep{kennicutt98, komugi05, schuster07, onodera10}
 typically scatter at least factor of $\sim 2$.
Recent studies \citep{kennicutt07, schruba10, onodera10, liu11} have found that the dispersion in the SK law
increases as smaller regions are sampled within the galaxy, which are unlikely to be explained by observational
 uncertainties.  The significant dispersion is attributed to differences in the evolutionary stage of the molecular
 clouds \citep{onodera10}, which may be a large scale manifestation of molecular cloud types proposed by \citet{kawamura09}.

An interesting test of this hypothesis is to see whether the dispersion of the SK law decreases significantly
 when star forming molecular clouds of comparable age are sampled.  An interacting galaxy system is an 
ideal target for this purpose.  Star formation is expected to be triggered by galaxy collision \citep{barnes96, saitoh09},
 and indeed interacting galaxies such as the Antennae \citep{whitmore99, wilson00, ueda12} and IIZw 096 \citep{inami10} 
are known to host exceptionally young stellar clusters in the overlapping region and tidal tails.  If the interaction 
occurs over a timescale that is less than a typical timescale of star formation or molecular cloud evolution, and 
recent enough so that there has been time for only one generation of star formation, then the galactic collision-induced 
star formation in various regions of the system are expected to be at a comparable evolutionary stage.  The interacting
galaxy pair Taffy I, is a rare laboratory which meets these requirements.

\subsection{Taffy I}
Taffy I (see figure 1) is an interacting system consisting of two galaxies (UGC12914/15) and the ``bridge'', 
an extended shock-induced synchrotron component connecting the two galaxies \citep[hereafter C93]{condon93}.
The distance is estimated to be 61 Mpc, from the systemic HI velocity of $4600\ \mathrm{km s^{-1}}$ (C93) and Hubble constant
 of $75\ \mathrm{km s^{-1} Mpc^{-1}}$.
The steepening of spectral index at the bridge and the simulated trajectory of the galaxies indicate a 
face-on collision which occurred only 20 Myr ago (C93). The galaxies are separating at a velocity of $\sim 450\ \mathrm{km\ s^{-1}}$ 
for the assumed Hubble constant (C93), and the relative inclination of the disks is $12^\circ$ \citep{giovanelli86}.  For a 
star forming disk of 7 kpc diameter (the separation between star forming regions E and G in UGC12915; see following 
sections and figure 1), the duration of the disk-disk collision is estimated to be $\sim 3$ Myr.
This is much less than the typical timescale of star formation in galactic disks
\citep[several $\times$ 10 Myr;][]{egusa04, egusa09} or the timescale of giant molecular clouds to evolve 
\citep[6-13 Myr;][]{kawamura09}.  Thus, the collision was instantaneous in terms of the star formation history
 of this system, and recent enough so that the triggered star formation would not have evolved for more than a generation.

Star formation is evident from H$\alpha$ and mid infrared observations, in both galactic disks and notably in the
 bridge region, where a giant HII region has been found \citep{bushouse87}.  The SFR of this system
 has not been well measured, however, because commonly used tracers fail to trace massive star formation reliable in this galaxy.

H$\alpha$ observation \citep{bushouse87} indicates a total SFR of $\sim 1.4\ M_\odot \mathrm{yr^{-1}}$ (see section 3.2)
.  However, the center of UGC\ 12915 was virtually undetected in H$\alpha$, indicating severe dust extinction.
\citet{gao03} used the 20cm continuum to derive the star formation rate, but cautioned that the 
20cm is contaminated largely in the bridge by the shock-induced synchrotron emission (C93) which is unrelated to the massive
 star formation.  Seven micron PAH features have been observed by \citet{jarrett99}, which returned 
a low global star formation rate of $ <\ 6\ M_\odot \mathrm{yr^{-1}}$ if we assume a conversion factor to
 SFR by \citet{roussel01}, but using a common conversion factor of mid-infrared PAH emission to SFRs is 
debatable in Taffy I, as the small dust particles traced at these wavelengths could have been destroyed by the 
strong shock induced by the large scale galactic collision \citep{braine03}.  The total infrared luminosity of L(IR)
 $ = 7\times 10^{10}\ L_\odot$ corresponds to a SFR of $12.1\ M_\odot \mathrm{yr^{-1}}$ using the conversion factor
by \citet{kennicutt98}.  This estimate is larger compared to those using other tracers, but the conversion factor assumes
 continuous and constant star formation over 10-100 Myr, which is questionable in these colliding galaxies.

Obtaining a more reliable estimation of the SFR within Taffy I and its age, requires 
an unbiased measure of massive star formation.  The Pa$\alpha$ emission line ($\lambda = 1.8751\ \mathrm{\mu m}$) is suited for this goal,
as the magnitude of its dust extinction is only 1/6 of that of H$\alpha$, and can trace massive star formation more 
directly than infrared dust emission.  Therefore, we have carried out a narrow-band imaging observation of the redshifted 
Pa$\alpha$ emission line.

\section{Observation and Data Reduction}
Narrow-band images by the $N191$ filter ($\lambda=1.91\mu m$, $\lambda/ \Delta \lambda =58$) covering the redshifted
 Pa$\alpha$ emission line and broad-band images in the $H$ and $K_s$ bands were obtained on 2010 October 9, 16, and 17,
 with Atacama Near Infrared camera \citep[ANIR;][]{motohara08} on the University of
 Tokyo Atacama 1.0m telescope \citep{minezaki10} installed at the summit of Co. Chajnantor (5640m altitude) in
 northern Chile, which is a part of the University of Tokyo Atacama Observatory Project \citep[PI: Yuzuru Yoshii;][]{yoshii10}. 
The extremely low precipitable water vapor (PWV $\sim$ 0.5mm) of the site enables observations of the Pa$\alpha$ 
emission line from the ground \citep{motohara10}.

Exposure times are 12420s, 2160s and 2160s for $N191$, $H$, and $K_s$, respectively. Standard reduction procedures of flat-fielding,
 self-sky creation and subtraction, and shift-and-add are carried out to obtain a final image for each band. Flux zero-point
are calculated using the magnitudes of 2MASS stars in the field of view.  The typical seeing is $\sim 1^{\prime \prime}$.
The 1.91$\mu$m continuum image is produced by the final images of the $H$ and $K_s$ band, and an emission-line image is created 
by subtracting the 1.91$\mu$m continuum image from the $N191$ image.

There are many atmospheric absorption features within the wavelength range of the $N191$ filter, which also vary temporally
 due to change of PWV. Therefore, we have used a newly developed method to restore Pa$\alpha$ line flux
 from an emission line image by estimating atmospheric absorption at the wavelength of the line, using model atmosphere by ATRAN
 software \citep{lord92}. This method is confirmed to reproduce the real flux within an accuracy of 10\%, by comparing the Pa$\alpha$
narrow-band images taken with HST/NICMOS (Tateuchi et al. in prep.).  The actual factor introduced by the ATRAN correction to our
Pa$\alpha$ data was 1.12, corresponding to PWV of $665.2\ \mathrm{\mu m}$.

\section{Results}
\subsection{Pa$\alpha$ Morphology}
The bottom panel of Figure 1 shows the continuum-subtracted Pa$\alpha$ image of Taffy I.  
Most star formation in this system as traced by Pa$\alpha$ is found to 
exist in discrete ``blobs'' which we denote as regions A through H.  The blobs are located in the centers of each galaxies
(regions B and F), and on their
spiral arms (regions A,C,E and G).  A notable blob is seen in the northern tail of UGC\ 12915 (region H), 
which is also bright in H$\alpha$ \citep{bushouse87}.  The giant HII region in the bridge (region D) is almost as bright as 
those at the galaxy centers.

\subsection{Star Formation Rate}
\citet{bushouse87} observed Taffy I in H$\alpha$, giving luminosities 
$L(\mathrm{H\alpha})=1.17\times 10^{41}$ and $L(\mathrm{H\alpha})=6.31\times 10^{40} \ \mathrm{erg\ s^{-1}}$,
corresponding to SFRs of 0.93 and 0.50 $\mathrm{M_\odot yr^{-1}}$ using the prescription of \citet{kennicutt98}
for UGC\ 12914 and UGC\ 12915, respectively.

We compare the observed Pa$\alpha$ and H$\alpha$ fluxes in each of the regions, assuming intrinsic line ratio of 
$\mathrm{Pa\alpha}/\mathrm{H\alpha}=0.12$ \citep[$\mathrm{T}=10^4 \ \mathrm{K}$ and case B recombination;][]{osterbrock89}
 and derive average dust extinction $A_V$ and extinction corrected Pa$\alpha$ flux (Table 1) using 
the \citet{rieke85} extinction law.
The extinction corrected Pa$\alpha$ luminosity $L$(Pa$\alpha$) and the assumed line ratios are
used to obtain the intrinsic H$\alpha$ luminosity, and then to SFR using the prescription by \citet{kennicutt98}.
\cite{bushouse87} assign an error of $\sim 20\%$ to the H$\alpha$ photometric flux.  Combined with the 10\% uncertainty in the
Pa$\alpha$ and error propagation in the calculations, the final error in the extinction corrected Pa$\alpha$ flux is estimated to be
$\sim 25\%$.  The flux for each of the blobs is measured using a circular aperture of $12^{\prime \prime}.6$ diameter,
 corresponding to a projected scale of 3.7 kpc.  Subtraction of diffuse emission (if any) is done using an
 annulus with a width of $5^{\prime \prime}$ around each aperture.

Derived values for each of the star forming blobs are listed in table \ref{blobs}.  The total star formation rate
of UGC\ 12914 and UGC\ 12915 (sum of constituent blobs) are $7.3$ and $14.9\ M_\odot \mathrm{yr^{-1}}$, totaling
$22.2 \ M_\odot \mathrm{yr^{-1}}$, an order of magnitude higher than
those derived from H$\alpha$ \citep{bushouse87}, or factor 4 higher than that
based on the $7 \mathrm{\mu m}$ PAH emission \citep{jarrett99, roussel01}.  It is also nearly two times higher than the
global SFR derived from the total IR luminosity.  Since the Pa$\alpha$ traces massive star formation more directly than the 
PAH or IR luminosity, it is possible that the SFR derived from these secondary tracers underestimate the true SFR because
of dust destruction \citep{braine03}, or due to the assumption of continuous star formation over 10-100 Myr in the case of the total IR luminosity.

\section{Discussion}
\subsection{Age}
Equivalent widths of the Pa$\alpha$ line are derived by dividing the Pa$\alpha$ image
with the $1.91 \mathrm{\mu m}$ continuum.
Approximate ages of the blobs are estimated from the equivalent widths using figure 5 of \citet{diaz08}
 which is based on Starburst99 \citep{leitherer99} with an instantaneous burst, and shown in Table 1.
The signal to noise ratio (SNR) of the Pa$\alpha$ blobs within the measured aperture ranged from 
SNR $\sim$7 to over 20, except for Region H in the faint tail of UGC12915, for which the SNR was 3.3.

The ages of all blobs except for region B (UGC\ 12914 center) and
 region D (the bridge) are $\sim 7$ Myr.  Region D is significantly younger, less than 3.5 Myr.  Region B is
 somewhat older, which is likely a contribution from an older population of stars as this is the center of the galaxy. 
 In any case, the estimated ages of the star forming blobs are less than 20 Myr, the estimated time since the 
collision of the two galaxies. The ages are surprisingly uniform considering the time since collision, but this may
 be explained by the Pa$\alpha$ observation sampling the brightest regions in the system which are forming stars
most actively at this point.

\subsection{Star Formation Efficiency and Schmidt-Kennicutt Law}
The total molecular gas mass of this system is estimated to be $M(\mathrm{H_2}) = 1.0\times 10^{10}\ \mathrm{M_\odot}$ 
(\citet{braine03}; Kaneko et al. submitted) using a CO-to-$\mathrm{H_2}$ conversion factor of 
$5.6\times 10^{19}\ \mathrm{cm^{-2} [K\ km\ s^{-1}]^{-1}}$ \citep{zhu07}.  From the global SFR estimate of
 $22.2\ M_\odot \mathrm{yr^{-1}}$, the average star formation efficiency (SFE=SFR/$M(\mathrm{H_2})$) in Taffy I
 is $2.2 \times 10^{-9} \mathrm{yr^{-1}}$, corresponding to a gas consumption timescale $\tau_\mathrm{gas}$ of $4.5\times 10^8$ yr.  
This gas consumption timescale is consistent with typical starburst galaxies \citep{kennicutt98}, and qualifies 
Taffy I as a starburst galaxy in contrast to indications from previous studies \citep{gao03}.

In order to derive the molecular gas mass and SFE of individual blobs, we use interferometric observations of
 the $\mathrm{^{12}CO}(J=1-0)$ line \citep{iono05}.  The angular resolution of the CO data is 
$7^{\prime \prime}.2 \times 5^{\prime \prime}.1$, sufficient to match the aperture ($12^{\prime \prime}.6$)
used to determine the SFRs and the equivalent widths of the Pa$\alpha$ emission.
  For the conversion factor, we use both the average $\mathrm{X_{CO}}$ and the values derived individually for
 UGC12914, UGC12915 and the bridge by \citet{zhu07} using a one-zone LVG analysis.  The derived gas masses are
 shown in Table \ref{blobs} for both cases.
The SFEs of the blobs are surprisingly uniform, with $\mathrm{SFE}=(7.2\pm 0.9) \times 10^{-9} \mathrm{yr^{-1}}$,
 corresponding to  $\tau_\mathrm{gas}=(1.4 \pm 0.2) \times 10^8$ yr, when the average conversion factor is used.
When different conversion factors are used for the galaxies and the bridge, the bridge has an exceptionally high
 SFE and short gas consumption timescale, with $\mathrm{SFE}=(1.4 \pm 0.1) \times 10^{-8} \mathrm{yr^{-1}}$ and 
$\tau_\mathrm{gas} =(7.1\pm 0.7) \times 10^7$ yr, whereas all the other regions have nearly constant SFE and $\tau_\mathrm{gas}$, of 
$\mathrm{SFE}=(4.9\pm 0.9) \times 10^{-9} \mathrm{yr^{-1}}$ and $\tau_\mathrm{gas}=(2.0 \pm 0.4) \times 10^8$ yr.  These values are all
 typical of circumnuclear starburst regions in nearby galaxies \citep{kennicutt98}.


The variation in the SFEs is also apparent from the comparison between molecular gas mass and SFR, shown in Figure \ref{schmidt}.
  Since the projected areas of the apertures are same for all the blobs, Figure \ref{schmidt} can also be regarded
as the SK law.  For the constant conversion factor, a best-fit to all the blobs results in
\begin{equation}
\log \Sigma_\mathrm{SFR}=(0.99 \pm 0.08)\log \Sigma_\mathrm{H2} - (8.13 \pm 0.12)
\end{equation}
and when using different conversion factors, all blobs excluding region D (bridge) results in 
\begin{equation}
\log \Sigma_\mathrm{SFR}=(0.95 \pm 0.14)\log \Sigma_\mathrm{H2} - (8.23 \pm 0.24)
\end{equation}
The observed power law index of $N=1.0$ is consistent with those typically observed using dense gas tracers, but not with
 with the typically observed $N>1$ using $\mathrm{^{12}CO}(J=1-0)$.

\subsection{Schmidt-Kennicutt Law and Dispersion}
The dispersion $\sigma$ of the correlation from the best fit is 0.058 dex (14\%) in case of using a constant conversion factor, or
0.096 dex (25\%) when using different conversion factors and excluding region D.
The $\sigma$ are comparable to or smaller than uncertainties in the SFR (25\%), so uncertainties in the measurement alone
 can account for the dispersion for all regions with comparable ages, and only the exceptionally young region D is offset
 significantly from this correlation.

The dispersion can be compared to those in normal galaxies which contain star forming regions at a variety of evolutionary
stages.  Although we use an aperture with a projected scale of 3.7 kpc, this is limited by the
CO resolution.  Inspection of the high resolution Pa$\alpha$ image reveals that
 even with this large aperture size, only one bright star forming complex is enclosed in each of the apertures.
The effect of averaging over star forming regions which work to decrease $\sigma$ \citep{liu11} would not be prominent in our 
aperture, so we compare the $\sigma$ in the literature at a common scale of $\sim 700$ pc, the typical size of Pa$\alpha$ blobs in
Taffy I.  \citet{liu11} derive $\sigma = 0.50$ and $0.27$ for M51 and NGC3521, respectively, at 700 pc resolution.  Similarly, 
we derive a dispersion of $\sigma = 0.43$ from the M33 dataset published by \cite{onodera10} at 500 pc resolution,
 and $\sigma = 0.32$ at 1 kpc resolution.  Given that our star forming blobs are distributed widely both within and between galaxies,
 the measured dispersion in Taffy I is unexpectedly small.

\subsection{Star Formation History and Evolution of Taffy I}
A stellar mass of a galaxy reflects its integrated star formation history, and therefore the
 ratio of current star formation rate to total stellar mass (the \textit{specific star formation rate}; sSFR)
indicates the intensity of current star formation relative to its average in the past.

The $K$ band magnitudes of UGC12914 ($9.32\ \mathrm{mag.}$) and UGC12915 ($9.83\ \mathrm{mag.}$; \cite{jarrett99}) can
be used to estimate their rotation velocities and hence their stellar mass based on the Tully-Fischer relation
\citep{bell01}, which give stellar mass estimates of $M_* = 7\times 10^{10}\ M_\odot$ and $M_* = 5\times 10^{10}\ M_\odot$
 for UGC12914 and UGC12915, respectively.  Using the SFR of $22.2\ M_\odot \mathrm{yr^{-1}}$,
 the sSFR of Taffy I is $1.9\times 10^{-10}\ \mathrm{yr^{-1}}$.  This is five times larger than the average of 
local ($0 \le z \le 0.2$) galaxies with similar stellar mass ($\sim 4\times 10^{-11}\ \mathrm{yr^{-1}}$) given by \citet{oliver10},
 indicating that Taffy I is now forming stars much more actively than its average past.  The ages of the blobs estimated in
 the previous section also indicate that the star forming regions studied here are experiencing their first burst of star formation. 
 Apparently, Taffy I has just evolved into a starburst system as a result of the collision which triggered massive star formation.






A scenario of star formation in Taffy I should explain the matched ages of the all regions except region D, and also the
apparent delay of star formation in region D.

\citet{braine03} and \citet{zhu07} suggest that the molecular gas in the bridge was ionized in the collision but
 cooled promptly, recombined and returned to the molecular state in several to 10 Myr \citep{bergin04}, delaying star formation
by that time.  This scenario requires the cooling time of the gas which created region D to be longer than the timescale of collision
($\sim$ 2 Myr or less: see section 1.1).  The cooling time of molecular gas at $10^{1-2} \ \mathrm{cm^{-3}}$ and $10^4$ K is $10^{4-6}$ years,
so the colliding gas may not have had enough time to be heated to delay molecular gas formation compared to other regions.
 Another possible scenario is that the molecular gas in region D was simply
 stretched as a result of tidal force, delaying aggregation of gas.  Assuming that UGC12915 is the 
dominant source of tidal force and has $M_* = 5\times 10^{10}\ M_\odot$, and using molecular mass of $1.2 \times 10^8 \ \mathrm{M_\odot}$
 and diameter of 1 kpc for Region D, the Roche limit is reached at 5 kpc.  The projected distance
of region D to UGC12915 is $\sim$ 4 kpc, so the tidal force may be strongly affecting region D.  It is plausible that only recently
the gas has been able to collapse locally within region D.

Apart from region D, the blobs can be regarded to be in a similar stage in their evolution from molecular gas to stars.  The SK law 
is consistent in this respect, as all the blobs except region D follow an exceptionally tight linear law, which is predicted
 if our working assumption that variations in the evolutionary stage of molecular clouds are responsible for the
 dispersion is correct.  Since the observed slope of $N = 1.0$ is typically observed with dense gas tracers, we speculate that the
 blobs in the same evolutionary stage have the same dense gas fraction, giving rise to a linear slope even when observed with 
$^{12}\mathrm{CO}(J=1-0)$ that traces the total molecular mass.  This can be tested by high resolution observations of dense 
molecular gas tracers, which should show that the dense gas fraction in these blobs are similar.


\section{Summary}
This paper presents ground-based narrow band imaging data of Pa$\alpha$ emission in Taffy I, an interacting galaxy system which
experienced a direct collision only 20 Myr ago.  The star formation rate of the system is larger than previous determinations using 
H$\alpha$, PAH emission, or the far infrared continuum.  The newly estimated star formation rate qualifies Taffy I as a starburst system.
Ages of the star forming blobs are estimated from equivalent widths of the Pa$\alpha$ emission, showing that virtually all
the detected star formation occurred after the collision, first in the centers of the galaxies and at the disks, then several Myr 
later at the bridge region.  This delay of star formation in the bridge region may be explained either by the time it took for
molecular clouds to cool after it ionized in the collision, or by self gravitational collapse hindered by tidal force of the 
parental galaxies.

Comparison with interferometric $^{12}\mathrm{CO}(J=1-0)$ data shows that the star forming blobs with the same age follow an
 exceptionally tight power law with no significant dispersion, whereas the young star forming region in the bridge is
offset from this correlation.  The slope of unity is unusual for the SK law using $^{12}\mathrm{CO}(J=1-0)$, but 
typical for cases when dense gas tracers are used to trace molecular gas.  These regions all show star formation 
efficiencies which are comparable to infrared
 luminous starbursts, and specific star formation rate are much higher than local galaxies with similar stellar mass.
We argue that Taffy I has just evolved into a starburst system, and the star forming blobs are at the
same evolutionary stage with a constant fraction of dense gas.
Our observations support a hypothesis where the dispersion in the SK law is largely due to variations in the evolutionary 
stage of the molecular clouds, and imply the possibility of defining different SK laws for different populations in a galaxy.

\acknowledgments
The authors acknowledge H. Bushouse for kindly providing the H$\alpha$ image, and S. Onodera for kindly providing the 
M33 dataset for deriving the dispersion in the SK law.  J.U. was supported by the Sasakawa Scientific Research Grant
from The Japan Science Society.  Operation of ANIR on the miniTAO 1m telescope is supported by Ministry of Education, Culture,
 Sports, Science and Technology of Japan, Grant-in-Aid for Scientific Research (20040003, 20041003, 21018003, 21018005, 21684006, and
22253002) from the JSPS, Research Grant for Universities, and Optical \& Near-Infrared Astronomy Inter-University Cooperation Program.  Part of this work has been supported by National Astronomical Observatory of Japan (NAOJ) Research Grant
for Universities.



{\it Facilities:} \facility{OVRO}, \facility{miniTAO (ANIR)}, \facility{ISO}, \facility{KPNO}.



\clearpage


\begin{onecolumn}

\begin{figure}
  \begin{center}
   \epsscale{1.3}\plottwo{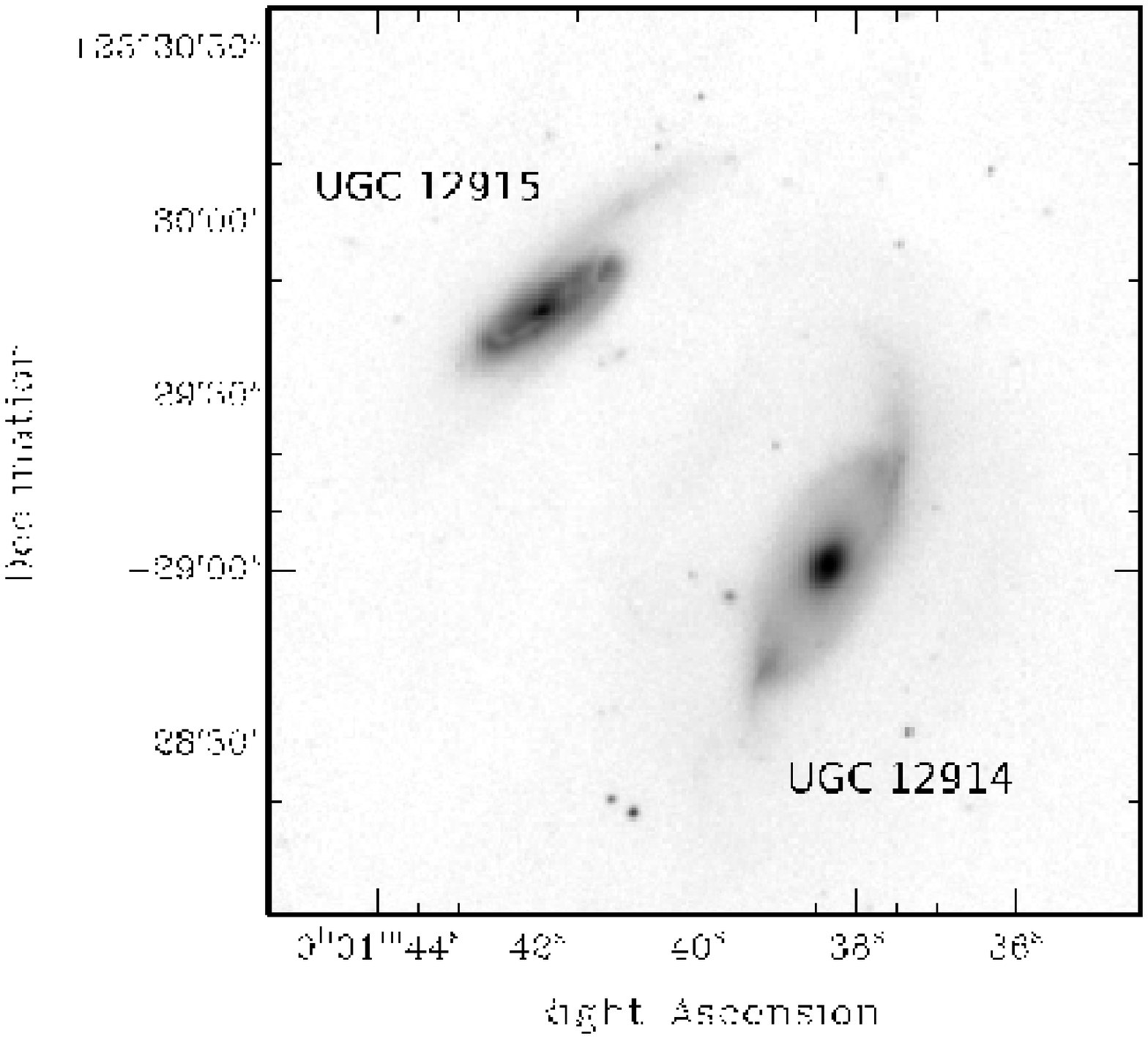}{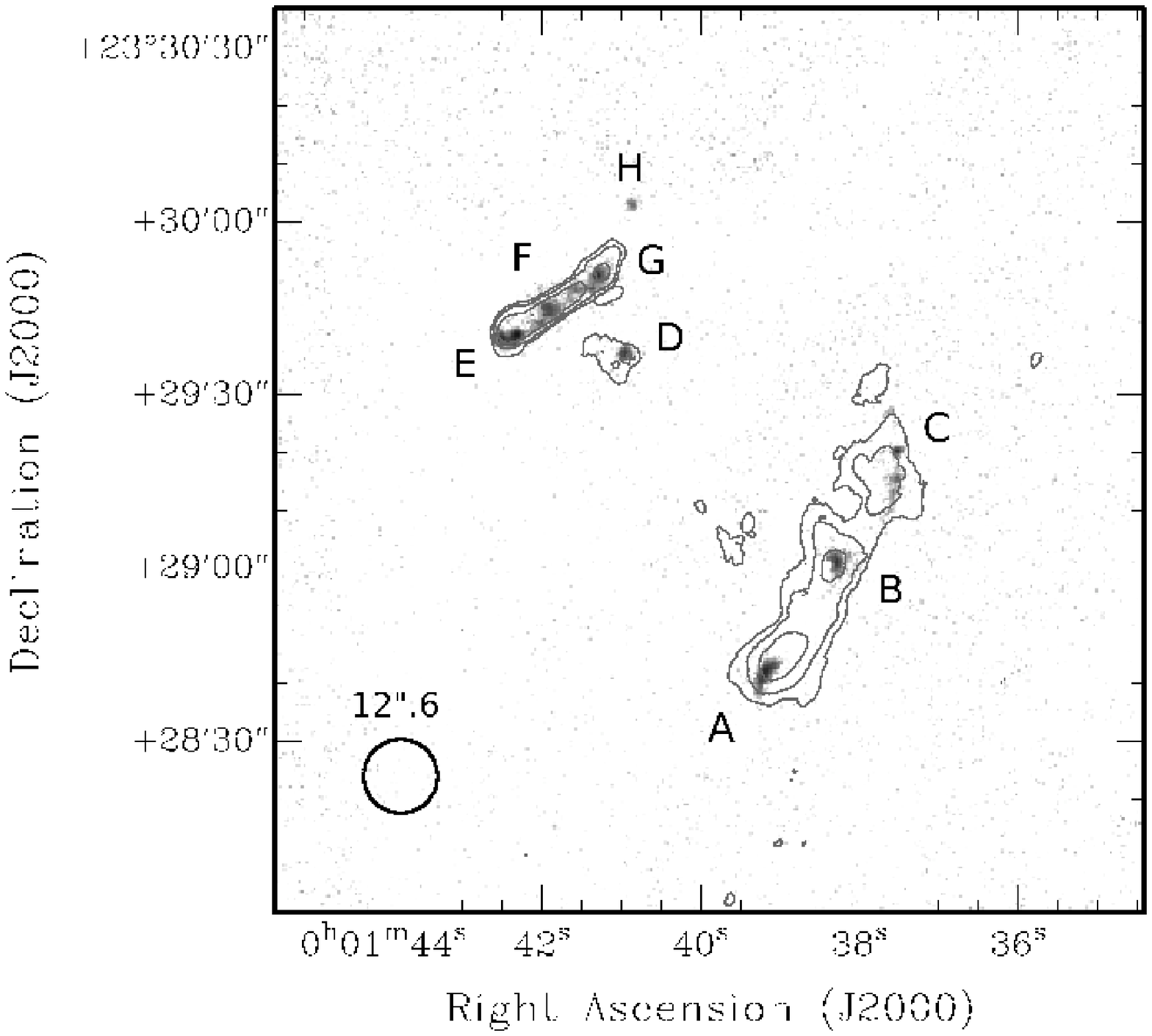}
  \end{center}
  \caption{Top: 1.91 $\mathrm{\mu m}$ continuum created from the $H$ and $K_s$ images.  Bottom: Pa$\alpha$ image
 with integrated CO intensity contours from \cite{iono05}.  Contour levels are at 
4, 8, 16 and 32 $\mathrm{Jy\ km\ s^{-1}}$.  Individual blobs are indicated by their alphabetical designations.  Circle in the lower left shows the aperture used for photometry.}
\label{vv254}
\end{figure}

\begin{figure}
  \begin{center}
  \epsscale{1}\plotone{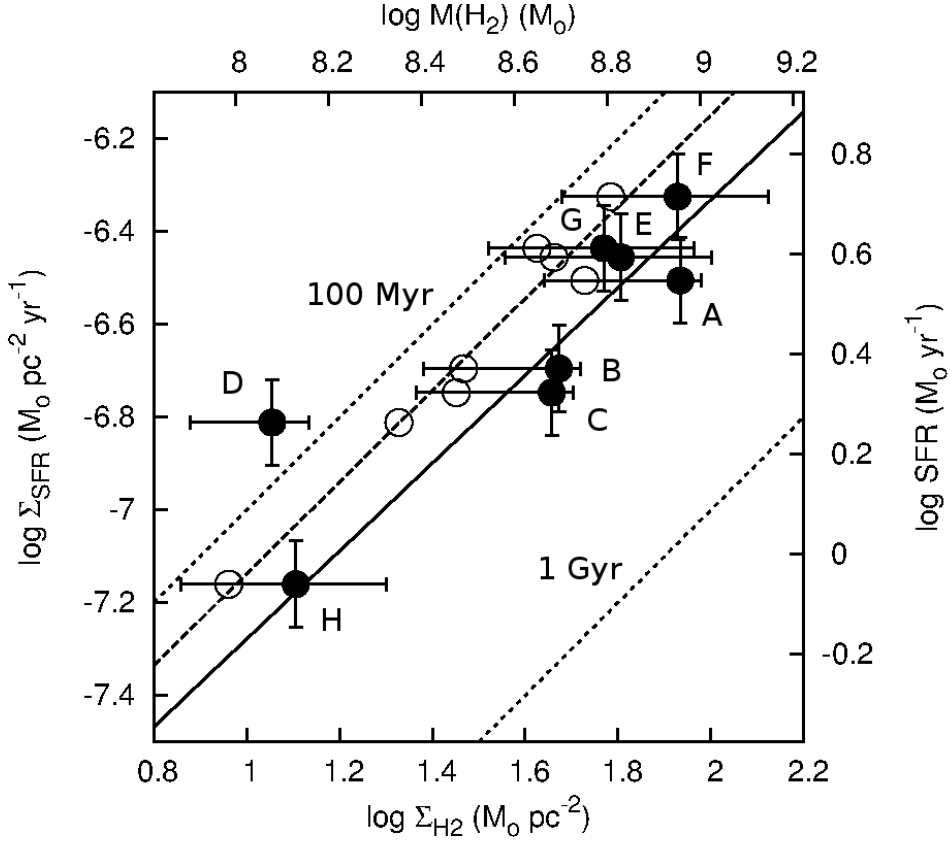}
  \end{center}
  \caption{Molecular gas mass and SFR in Pa$\alpha$ blobs.  Open circles use gas mass calculated using a constant
 $\mathrm{X_{CO}}$ (column 3 in table \ref{blobs}), and filled circles use varying $\mathrm{X_{CO}}$ (column 4 in table \ref{blobs}).
Uncertainties in the $\mathrm{X_{CO}}$ from \citet{zhu07} are used for the x-axis errorbars, 
and 10\% for the y-axis (see section 2).
Solid and long dashed lines are the best fit power law (equation 2 and 3) for constant and varying $\mathrm{X_{CO}}$, respectively.
The short dashed lines show constant gas consumption timescales of $10^8$ and $10^9$ yr.}
\label{schmidt}
\end{figure}

\end{onecolumn}

\clearpage

\begin{table}
\tabletypesize{\scriptsize}
  \caption{Properties of Pa$\alpha$ blobs}
  \label{blobs}
  \begin{center}
    \begin{tabular}{llcccccccc}
      \tableline\tableline
      Name & Region & $\mathrm{M(H_2)}$ & $\mathrm{M(H_2)_{var}}$ & $A_V$ & $\mathrm{L(Pa\alpha)}$ & $\mathrm{EW(Pa\alpha)}$ & SFR & $\tau_\mathrm{gas}$  & Age \\ 
           &        & $\mathrm{10^8\ M_\odot}$ & $\mathrm{10^8\ M_\odot}$  & mag.  &  $\mathrm{10^{40}\ erg\ s^{-1}}$  & $\log \mathrm{\AA}$ 
 & $\mathrm{M_\odot \ yr^{-1}}$ & $100$Myr & Myr \\ \tableline
    A & U12914 S   &  5.65 & 9.07 & 6.44 & 4.86 (1.72) & 1.86 & 3.30 (1.17) & 2.8  & 6.5 \\
    B & U12914 C   &  3.10 & 4.97 & 5.56 & 3.14 (1.28) & 0.95 & 2.13 (0.87) & 2.3  & $>$ 7.9 \\
    C & U12914 N   &  2.99 & 4.80 & 5.49 & 2.79 (1.15) & 1.66 & 1.89 (0.78) & 2.5  & 6.8 \\
    D & bridge     &  2.25 & 1.20 & 6.20 & 2.39 (0.88) & 3.85 & 1.63 (0.60) & 0.7  & $<$ 3.5 \\
    E & U12915 S   &  4.85 & 6.76 & 6.26 & 5.46 (1.99) & 1.54 & 3.70 (1.35) & 1.8  & 6.9 \\
    F & U12915 C   &  6.43 & 8.96 & 7.22 & 0.74 (0.23) & 1.35 & 5.00 (1.56) & 1.8  & 7.2 \\
    G & U12915 N   &  4.46 & 6.21 & 7.21 & 0.58 (0.18) & 1.57 & 3.87 (1.21) & 1.6  & 6.9 \\
    H & U12915 tail&  0.97 & 1.35 & 5.54 & 1.08 (0.44) & 1.82 & 0.73 (0.30) & 1.9  & 6.6 \\
\tableline
    \end{tabular}
\tablenotetext{}{col. (1)(2) Names of Pa$\alpha$ blobs.  Suffix are N for north, S for south, and C for center.}
\tablenotetext{}{col. (3) Molecular mass in $12^{\prime \prime}.6$ aperture, using data from \citet{iono05}, for a 
distance of 61 Mpc and $\mathrm{X_{CO}}=5.6 \times 10^{19}\ \mathrm{cm^{-2} [K\ km\ s^{-1}]^{-1}}$.}
\tablenotetext{}{col. (4) Same as column 3, for varying $\mathrm{X_{CO}}$, using $9.0\times 10^{19}$, $7.8\times 10^{19}$ and 
$3.0\times 10^{19}\ \mathrm{cm^{-2} [K\ km\ s^{-1}]^{-1}}$ for UGC12914, UGC12915, and the bridge, respectively \citep{zhu07}.}
\tablenotetext{}{col. (5) V band extinction, measured from the Pa$\alpha$/H$\alpha$ ratio.}
\tablenotetext{}{col. (6) Extinction corrected Pa$\alpha$ luminosity in the $12^{\prime \prime}.6$ aperture.  Values in 
parenthesis are for no extinction correction applied.}
\tablenotetext{}{col. (7) Equivalent width of Pa$\alpha$.}
\tablenotetext{}{col. (8) Star formation rate obtained using column 6 and prescription of \citet{kennicutt98}.  Meaning of parenthesis same as column 6.}
\tablenotetext{}{col. (9) Gas consumption timescale, from column 4 divided by column 8.}
\tablenotetext{}{col. (10) Age of the region, estimated from column 7.}
  \end{center}
  \end{table}

\end{document}